\documentclass[prd,a4paper,nofootinbib,
showpacs
]{revtex4}

\usepackage{graphicx}

\def\eq#1{{eq.~(\ref{#1})}}
\def\eqs#1#2{{eqs.~(\ref{#1})--(\ref{#2})}}

\def\vev#1{\left\langle #1\right\rangle}

\def\Re{\mbox{Re}\,}
\def\Tr{\mbox{Tr}\,}
\def\det{\mbox{det}\,}

\def\ltap{\ \raisebox{-.4ex}{\rlap{$\sim$}} \raisebox{.4ex}{$<$}\ }

\def\hbar{\hspace{0pt}\raisebox{1pt}{$-$} \hspace{-7pt} h}

\def\5{\overline 5}

\newcommand{\be}{\begin{equation}}
\newcommand{\ee}{\end{equation}}
\newcommand{\bea}{\begin{eqnarray}}
\newcommand{\eea}{\end{eqnarray}}
\newcommand{\nn}{\nonumber}

\begin{document}
\title[Little Flavon]{Fermion Masses and Mixings in the Little Flavon Model
}
\date{September 12, 2003}
\author{F.~Bazzocchi$^\dag$}
\author{S.~Bertolini$^\dag$}
\author{M.~Fabbrichesi$^\dag$}
\author{M.~Piai$^\ddag$}
\affiliation{$^\dag$INFN, Sezione di Trieste and\\
Scuola Internazionale Superiore di Studi Avanzati\\
via Beirut 4, I-34014 Trieste, Italy\\
$^\ddag$Department of Physics, Sloane Physics Laboratory\\
University of Yale, 217 Prospect Street\\
New Haven CT 06520-8120, USA}
\begin{abstract}

\noindent
We present a complete analysis of the fermion masses and mixing matrices in the
framework of the little flavon model. In this model textures are generated by
coupling the fermions to scalar fields, the little flavons, that are pseudo-Goldstone bosons of the breaking of a global $SU(6)$ symmetry.
The Yukawa couplings arise from the  vacuum expectation values of the flavon fields, their sizes
 controlled by a potential  \`a la Coleman-Weinberg.
Quark and lepton mass hierarchies and mixing angles are accomodated within the effective approach in a natural manner.
\end{abstract}

\maketitle
%

\vskip1.5em
\section{Framework}

Fermion masses and mixing angles  can in principle be explained by patterns of vacuum expectation values of one or more  fields (the \textit{flavons}) in spontaneously broken horizontal
symmetries~\cite{flavormodels-old,flavormodels-new}.
The hierarchies in these expectation values---which must  be present in order to explain the experimental data---require however some mechanism in order to be stabilized without resorting to fine tuning of the parameters of the theory.

A possible framework solving this problem of fine tuning of the parameters---and providing an explicit effective potential from which the vacuum expectation values can be derived---has been introduced in ref.~\cite{BBFP} where the spontaneous breaking of the horizontal symmetry
is driven by the vacuum expectation values (VEV's) of flavon fields originally arising
as pseudo-Goldstone bosons of a larger group. The mechanism used in this model to stabilize
the breaking scale against one-loop quadratically divergent radiative correction is that of
\textit{collective} breaking already employed in the little Higgs models
in the context of electroweak symmetry breaking~\cite{littlehiggs}.

In the model, the flavons are the pseudo-Goldstone bosons of the breaking of a flavor
symmetry group $SU(6)$ down to $Sp(6)$.
Fourteen of the generators of $SU(6)$ are broken
giving 14 (real) Goldstone bosons that can be written as a single field
\be
\Sigma = \exp \left( i \Pi/f \right) \label{sigmone}
\Sigma_0 \, .
\ee
They represent  fluctuations around the (anti-symmetric)
vacuum expectation value
\be
\Sigma_0 \equiv \langle \Sigma \rangle =
\left( \begin{array}{cc} 0 & -I \\ I & 0 \end{array} \right) \, .  \label{vacuum}
\ee

At the same time, four subgroups
$[SU(2)\times U(1)]^2$ defined in ref.~\cite{BBFP}
are gauged. The gauge transformations explicitly break the global $SU(6)$
symmetry thus giving mass to the pseudo Goldstone bosons.
In the low-energy limit,
we can write the pseudo-Goldstone boson matrix as
\be
\Pi = \left( \begin{array}{cccccc} 0& 0& \phi_1^+ & 0 & s & \phi_2^0 \\
                                                    0& 0& \phi_1^0 &  -s & 0 & \phi_2^- \\
                                                     \phi_1^- & \phi_1^{0*} &0 & -\phi_2^0 & -\phi_2^- &0 \\
                                                     0 & -s^* & -\phi_2^{0*} & 0 & 0 & \phi_1^{-}\\
                                                     s^* & 0 & -\phi_2^+ & 0 & 0 & \phi_1^{0*}\\
                                                     \phi_2^{0*} & \phi_2^+ & 0 & \phi_1^+ & \phi_1^0 & 0
                                                     \end{array} \right) \, .
\ee
The singlet field becomes massive and has no expectation value in the vacuum configuration we will use; it is therefore effectively decoupled from the theory. The two doublets are our \textit{little flavons}:
\be
\phi_1 = { \phi_1^+  \choose \phi_1^0} \quad \mbox{and}\quad
\phi_2 = {\phi_2^0 \choose \phi_2^-} \, ,
\ee
 that are $SU(2)$-doublets  with $U(1)$ charges respectively $1/2$ and $-1/2$,
 and one $SU(2)$- and $U(1)$- singlet $s$.

The Coleman-Weinberg
effective potential~\cite{coleman-weinberg}, induced by $SU(6)$ non-symmetric
gauge and Yukawa interactions has been discussed in \cite{BBFP}
and takes the form:
\be
 \mu_1^2 \phi_1^\dag \phi_1 + \mu^2_2 \phi_2^\dag \phi_2+
\lambda_1 (\phi_1^\dag \phi_1)^2 + \lambda_2 (\phi_2^\dag \phi_2)^2
 +   \lambda_3 (\phi_1^\dag \phi_1)(\phi_2^\dag \phi_2)
+ \lambda_4 \,  |\tilde{\phi_2}^\dag \phi_1|^2  \, .
\label{CWpot}
\ee
An explicit analysis of this potential shows that at the leading order the
quadratically divergent part contains only the term
$\lambda_4= g_1^2 g_2^2/(g_1^2 + g_2^2) $  and there are
no mass terms for the flavon doublets.
This result is in agreement with the general features of the little Higgs models.

The logarithmitcally divergent part generates the other quartic couplings
in \eq{CWpot} with a typical size given by~\cite{BBFP}
\bea
\lambda_{1,2}& \simeq& \frac{\lambda_4^2}{64 \pi^2}
\log \frac{\Lambda^2}{M_\phi^2}  \ltap 10^{-2}\, ,
 \nn \\
\lambda_{1,2,3} & \simeq &  c_{1,2,3} \frac{3g^4}{64 \pi^2}
\log \frac{\Lambda^2}{M_V^2}  \ltap 10^{-2} \, ,  \label{L1L2loopL4}
\eea
where $c_i$ are numerical coefficients, related to the  expansion of the $\Sigma$, and $M_V \simeq f$ is the mass of the massive gauge bosons. $\Lambda = 4 \pi f$ is the model cut-off scale that we take around 100 TeV.
Explicit $SU(6)$ breaking in the Yukawa interactions leads to quadratically
divergent contributions to flavon mass terms of the form
\be
\mu_{1, 2}^2 \simeq  - c^{(1,2)}_{n} \, \eta_{n}\ \frac{\Lambda^2}{16\pi^2}
\simeq - c^{(1,2)}_{n}\,  \eta_{n}\ f^2 \, ,
\label{mu12eta}
\ee
where   $c^{(1,2)}_{n}$ are coefficients of order unity and
$\eta_i$ are Yukawa couplings in the right-handed
neutrino sector of order $\ltap 10^{-2}$. Their presence induces
negative $\mu_{1, 2}^2$ thus triggering the spontaneous breaking
of the residual $SU(2)\times U(1)$ gauge flavor symmetry.
On the other hand, the size of the induced quartic  flavon couplings remains small enough not to significantly affect their mass spectrum~\cite{BBFP}.

A residual $U(1)$ global symmetry, acting with opposite charge
on $\phi_i$ and $\tilde{\phi}_i$ fields, forbids the generation of
mixed terms proportional to $\tilde \phi_1^\dag \phi_2$. Therefore
the vacuum which completely breaks
the $SU(2)\times U(1)$ gauge flavor symmetry can be parametrized as
\be
\langle \phi_1 \rangle = { 0  \choose v_1} \quad \quad \langle \phi_2
\rangle = {0 \choose v_2}\, ,
\label{vacuum2}
\ee
where $v_1$ and $v_2$ are real.

Albeit preserved by the vacuum, the residual global $U(1)_P$ symmetry,
is explicitly broken by the lepton Yukawa sector~\cite{BBFP} in a way that
maintains to a high accuracy the vacuum structure in \eq{vacuum2}.
The scalar potential of the model contains in addition to \eq{CWpot}
the standard Higgs potential and flavon-Higgs mixing terms, which, as
discussed in ref. \cite{BBFP}, neither destabilize the standard
electroweak vacuum nor the flavon vacuum considered above.

Let us stress that the only  large couplings  present in the model---with the exception
of the top Yukawa coupling on which we comment in sect.\ II.B---are
the flavor gauge couplings which are taken to be $O(1)$ to make the flavor gauge
bosons sufficiently heavy after spontaneous breaking of the symmetry. As we have seen, the scalar sector of the theory is protected from large loop corrections induced by these
gauge interactions by a little-Higgs-like mechanism. This feature, together with the presence of small effective
Yukawa and scalar couplings (due to vacuum induced suppression factors),
makes the model highly
stable against radiative corrections from all sectors.

All couplings in the potential and overall fermion scales take natural values
(not smaller than $10^{-2}$), the only exception being the overall scale of the neutrino mass matrix which implies a Yukawa coupling of
$O(10^{-4})$ because of the smallness of the see-saw scale in the model.
The large mass hierarchies, present in the quark
and lepton sectors only come from the texture generated
by the spontaneous breaking of the flavor symmetry.

\section{Textures generation}

As discussed in ref.~\cite{BBFP}, the spontaneous breaking of
the global $SU(6)\rightarrow Sp(6)$ (approximate) symmetries
leads to the breaking of the gauged $[SU(2) \times U(1)]^2$ subgroup to $SU(2) \times U(1)$.
Fermions of different families transform according to the $SU(2)_F\times U(1)_F$ gauge flavor symmetry, labeled
by the index $F$ in order to distinguish it from the standard electroweak group.
In the following, all Greek
indices are related to the flavor group while Latin indices refer to the
electroweak group.

Textures in the mass matrices of fermions are generated by coupling the
flavon fields to the fermions, after the spontaneous breaking of the flavor symmetry.
The model does not explain the overall scales of the fermion masses, that
have to be put in by hand; it explains the hierarchy among families that exists after that scale has been fixed.

The effective lagrangians are rather cumbersome because many
different couplings are allowed by the flavor symmetry.
The little flavon fields enter as components of the pseudo-Goldstone field $\Sigma$
introduced in \eq{sigmone} of the effective non-linear sigma model.
In \cite{BBFP} we have shown how the required textures in the leptonic
sector arise naturally from the vacuum structure, the precise
fit of masses and mixings being determined by detailed values
of unknown Yukawa couplings which differ at most by factors of order one.
We complete here the discussion by including also the quark sector and
 presenting a global fit of quark and lepton masses and mixing angles
that arises without large hierarchies or fine tuning of the Yukawa coefficients in sect.\ III.

\subsection{Generalities}

After electroweak symmetry breaking, the effective lagrangian
contains the following mass terms for fermions:
\bea
{\cal L}^{(m)}\,=\,-\,
\bar{\psi}_R^{(i)}\,M^{(i)} \psi_L^{(i)}\,
-\,\frac{1}{2}\,\chi^{T}\,C\, M^{(n)}\chi\,
      +\,{ H.c.}\,,
\eea
where $\psi_{L,R}^{(i)}$ are chiral fields, $M^{(i)}$ are
$3\,\times\,3$ matrices, $i=u,d,l$, $M^{(n)}$ is
a $6\,\times\,6$ symmetric matrix,
$\chi\,=\,(\nu_{L},C\nu^{\ast}_{R})^T$ and
flavor indeces are understood.
$C$ is the charge conjugation matrix.

The neutrino mass matrix can be written in  $3\,\times\,3$ block form
as:
\bea
M^{(n)}\,=\,\left(\begin{array}{cc}
m_L & m_D^T \cr
m_D & m_R \end{array}\right)\,.
\eea
In the present case, $m_L = 0$ and  the scale of $m_R$ (whose generation does not involve neither electroweak nor flavor symmetry breaking at leading order)  is of order  $ f$ and therefore much larger than that of $m_D$. In the spirit of
effective field theory, one can approximately block-diagonalize
$M^{(n)}$, decouple three heavy states which are predominantly
standard model singlets, and write the Majorana mass term for the
light states as
\bea
{\cal L}^{(m)}\,=\,-\frac{1}{2}\,\nu_L^{T}\,
   C\,M^{(\nu)}\,\nu_L\, +\,{ H.c.}\,,
\eea
where now the (symmetric) Majorana mass matrix for the light fields (with some abuse of notation, we identify the light fields with the left-handed components) is $M^{(\nu)}\,=\,-m_D^T\,m_R^{-1}\,m_D$.

All matrices are non-diagonal in flavor space. One can
diagonalize them with appropriate bi-unitary trasformations,
\bea
\mbox{diag}\,M^{(i)}\,&=&\,R^{(i)\,\dagger}\,M^{(i)}\,L^{(i)}\,,\\
\mbox{diag}\,M^{(\nu)}\,&=&\,L^{(\nu)\, T}\,M^{(\nu)}\,L^{(\nu)}\,,
\eea
where  $L^{(i)}$, $R^{(i)}$ ($i=u,d,e$) and $L^{(\nu)}$ are $3\times3$ matrices in flavor space. With these definitions one finds that the
mixing matrices appearing in the charge-current interactions
according to the standard notation are given by
\bea
V_{CKM}\,&=&\,L^{(u)\,\dagger}\,L^{(d)}\label{ckm}\,,\\
V_{PMNS}\,&=&\,L^{(l)\,\dagger}\, L^{(\nu)} \,,
\label{pmns}
\eea
for quarks and leptons, respectively. We use the standard definitions
of the mixing matrices, in which one writes the down-type quark
(neutrino) flavor eigenstates  $d^{\prime}$ ($\nu^{\prime}$)
in terms of the mass eigenstates $d$ ($\nu$)---in the basis in which up-type
quarks (charged leptons) are diagonal--- as
\bea
d^{\prime} & = & V_{CKM} \, d\,,\\
\nu^{\prime} & = & V_{PMNS} \, \nu\,.
\eea

The standard parameterization of the Cabibbo-Kobayashi-Maskawa (CKM) matrix in terms
of three mixing angles $\theta_{12}$, $\theta_{13}$ and
$\theta_{23}$ and one  phase $\delta$ reads:
\bea
V_{CKM}\,=\,\left(\begin{array}{ccc}
c_{12}c_{13} & s_{12}c_{13} & s_{13}e^{-i\delta}\\
 - s_{12}c_{23} - c_{12}s_{23}s_{13}e^{i\delta} &
c_{12}c_{23} - s_{12}s_{23}s_{13}e^{i\delta} & s_{23}c_{13}\\
s_{12}s_{23} - c_{12}c_{23}s_{13}e^{i\delta}
& -c_{12}s_{23} - s_{12}c_{23}s_{13}e^{i\delta}
& c_{23}c_{13}\\
\end{array} \right)\,,
\eea
where $c_{ij}=\cos\theta_{ij}$ and $s_{ij}=\sin\theta_{ij}$.
An analogous expression is valid for the Pontecorvo-Maki-Nakagawa-Sakata (PMNS) matrix,
neglecting the flavor-diagonal Majorana phases.

\subsection{Quarks}

Quarks are characterized by small mixing angles. In this respect
it is natural to consider them as singlets under non-abelian
flavor symmetries. We take all standard model quarks---left-handed
electroweak doublet components as well as right-handed electroweak
singlets---to be singlets under $SU(2)_F$ while being charged
under $U(1)_{F}$. Textures generated by abelian symmetries have
been widely discussed in the literature~(see for instance
\cite{abelian} and references therein). Here we embed this ansatz in the little
flavon framework paying attention to the issue of the stability of
the flavon potential, while avoiding the large hierarchies among
the Yukawa couplings which are present in the standard model. A
possible charge assignment is summarized in Table I.
\begin{table}[ht]
\begin{center}
\caption{Summary of the charges of quarks and flavon fields ($\alpha= 2,3$)
under the horizontal flavor groups $SU(2)_F$ and $U(1)_F$.
$Q_{iL}$ stands for the electrweak
left-handed doublets. $q$ is an arbitrary charge that is not determined.}
\label{fields1}
\vspace{0.2cm}
\begin{tabular}{|c|c|c|}
\hline
&\quad\quad $U(1)_F$ \quad\quad &\quad $SU(2)_F$ \quad \cr
\hline
$Q_{1L}$ &   $q+3$ & 1 \cr
$Q_{2L}$  & $q+2$ & 1 \cr
$Q_{3L}$  & $q$ & 1 \cr
$u_{R}$  & $q-3$ & 1  \cr
$c_{R}$  & $q-1$ & 1  \cr
$t_{R}$  & $q$ & 1  \cr
$d_{R}$  & $q-4$ & 1  \cr
$s_{R}$  & $q-2$ & 1  \cr
$b_{R}$  & $q-2$ & 1  \\
\hline
$\Sigma_{\alpha-1\, 6} = (- i/f\ \phi_1 + ...)_{\alpha-1}$ &1/2 & 2 \\
$\Sigma_{\alpha-1\, 3} = (+ i/f\ \phi_2 + ...)_{\alpha-1}$ &$-1/2$ & 2 \\
$\Sigma_{3\, 2+\alpha} = (- i/f\ \phi_1^* + ...)_{\alpha-1}$ & $-1/2$&  $2^*$ \\
$\Sigma_{6\, 2+\alpha} = (- i/f\ \phi_2^* + ...)_{\alpha-1}$ &1/2 &  $2^*$ \\
\hline
\end{tabular}
\end{center}
\end{table}

Given the charges in Table I,  we find for the up quarks the following
effective Yukawa lagrangian
\bea
-\mathcal{L}_{u} &=&
\lambda_{31}\overline{t_{R}}\big(\Sigma_{\alpha-1\,3}\Sigma_{3\,2+\alpha}\big)^3\tilde{H}^\dag Q_{1L} +\lambda_{32}\overline{t_{R}}\big(\Sigma_{\alpha-1\,3}\Sigma_{3\,2+\alpha}\big)^2\tilde{H}^\dag Q_{2L} \nn \\
&+&\overline{t_{R}}(\lambda_{33} + \lambda_{33}'\Sigma_{\alpha-1\,6}\Sigma_{3\,2+\alpha}+ \lambda_{33}''\Sigma_{\alpha-1\,3}\Sigma_{6\,2+\alpha}\big)\tilde{H}^\dag Q_{3L}\nn \\
&+ &
\lambda_{21}\overline{c_{R}}\big(\Sigma_{\alpha-1\,3}\Sigma_{3\,2+\alpha}\big)^4\tilde{H}^\dag Q_{1L} +\lambda_{22}\overline{c_{R}}\big(\Sigma_{\alpha-1\,3}\Sigma_{3\,2+\alpha}\big)^3 \tilde{H}^\dag Q_{2L} \nn \\
&+& \lambda_{23}\overline{c_{R}}\big(\Sigma_{\alpha-1\,3}\Sigma_{3\,2+\alpha}\big) \tilde{H}^\dag Q_{3L}\nn \\
&+ &
\lambda_{11}\overline{u_{R}}\big(\Sigma_{\alpha-1\,3}\Sigma_{3\,2+\alpha}\big)^6\tilde{H}^\dag Q_{1L} +\lambda_{12}\overline{u_{R}}\big(\Sigma_{\alpha-1\,3}\Sigma_{3\,2+\alpha}\big)^5\tilde{H}^\dag Q_{2L}\nn \\
&+& \lambda_{13}\overline{u_{R}}\big(\Sigma_{\alpha-1\,3}\Sigma_{3\,2+\alpha}\big)^3 \tilde{H}^\dag Q_{3L}\, +\, H.c.
\label{yukup}
\eea
as well as
\bea
-\mathcal{L}_{d} &=&
\tilde{\lambda}_{31}\overline{b_{R}}\big(\Sigma_{\alpha-1\,3}\Sigma_{3\,2+\alpha}\big)^5 H^\dag Q_{1L} +\tilde{\lambda}_{32}\overline{b_{R}}\big(\Sigma_{\alpha-1\,3}\Sigma_{3\,2+\alpha}\big)^4 H^\dag Q_{2L} \nn \\
&+ &
\tilde{\lambda}_{33}\overline{b_{R}}\big(\Sigma_{\alpha-1\,3}\Sigma_{3\,2+\alpha}\big)^2 H^\dag Q_{3L}\nn \\
&+ &
\tilde{\lambda}_{21}\overline{s_{R}}\big(\Sigma_{\alpha-1\,3}\Sigma_{3\,2+\alpha}\big)^5 H^\dag Q_{1L} +\tilde{\lambda}_{22}\overline{s_{R}}\big(\Sigma_{\alpha-1\,3}\Sigma_{3\,2+\alpha}\big)^4 H^\dag Q_{2L} \nn \\
&+& \tilde{\lambda}_{23}\overline{s_{R}}\big(\Sigma_{\alpha-1\,3}\Sigma_{3\,2+\alpha}\big)^2H^\dag Q_{3L}\nn \\
&+ &
\tilde{\lambda}_{11}\overline{d_{R}}\big(\Sigma_{\alpha-1\,3}\Sigma_{3\,2+\alpha}\big)^7 H^\dag Q_{1L}
+\tilde{\lambda}_{12}\overline{d_{R}}\big(\Sigma_{\alpha-1\,3}\Sigma_{3\,2+\alpha}\big)^6 H^\dag Q_{2L}\nn \\
&+& \tilde{\lambda}_{13}\overline{d_{R}}\big(\Sigma_{\alpha-1\,3}\Sigma_{3\,2+\alpha}\big)^4 H^\dag Q_{3L}\, +\, H.c.
\label{yukdown}
\eea
for the down quarks.

Notice that even though the $t$ quark has a large Yukawa coupling that could introduce a
potentially destabilizing term  in the flavon effective potential,
the contribution to the flavon mass
terms of $t$-quark loops induced by the couplings in \eq{yukup}
\be
\mu_{1, 2}^2 \simeq
- \Re\left(\lambda_{33}^* (\lambda_{33}',\ \lambda_{33}'')\right) \,
\frac{\vev{h_0}^2}{f^2} \frac{\Lambda^2}{16\pi^2}
\label{mu12top}
\ee
remains negligible compared to \eq{mu12eta}.

\subsection{Leptons}

At variance with the quark sector, lepton flavor mixings differ 
by the presence of large angles, as implied by the neutrino oscillation data.
The maximal atmosferic neutrino oscillation
together with the hierarchy structure in the charged lepton mass spectrum,
suggest that the leptons of the second and third family may belong to
flavor doublets.
In the following we take
the weak electron doublet ${l}_{e L}$ to be an $SU(2)_{F}$
singlet charged under $U(1)_{F}$, while the standard model doublets ${l}_{\mu,\tau L}$ are members of a doublet in flavor space.
Right-handed charged leptons are assumed to follow a similar structure.

In order to have a \textit{see-saw}-like  mechanism~\cite{seesaw},
we introduce three right-handed neutrinos $\nu^{i}_{R}$  which are $SU(2)_{F}$ singlets.
This choice allows us to take right-handed neutrino mass entries at the scale  $M\sim f$.
Table II summarizes the charge assignments.
\begin{table}[ht]
\begin{center}
\caption{Summary of the charges of all leptons under the horizontal flavor groups $SU(2)_F$ and $U(1)_F$.}
\label{fields2}
\vspace{0.2cm}
\begin{tabular}{|c|c|c|}
\hline
&\quad\quad $U(1)_F$ \quad\quad &\quad $SU(2)_F$ \quad \cr
\hline
${l}_{e L}$ &   $-2$ & 1 \cr
$e_R$  & $2$ & 1 \cr
${L}_L = ( {l_{\mu}}\, ,\,{l_{\tau}})_L$  & $1/2$ & 2  \cr
$E_R = (\mu \,,\,\tau)_R$  & $1/2$ & 2  \cr
$\nu_{1R}$  & $1$ & 1  \cr
$\nu_{2R}$  & $-1$ & 1  \cr
$\nu_{3R}$  & $0$ & 1  \\[1ex]
\hline
\end{tabular}
\end{center}
\end{table}

The neutrino lagrangian is obtained after integrating out
the three right-handed neutrinos  and, at the leading order
in the  right-handed neutrino mass and in number of $\Sigma$ fields, is given by
(see \cite{BBFP}):
\bea
-2\, \mathcal{L}_{\nu} & = &
\frac{(\overline{{l}^c_{1L} }\ \tilde H^*)(\tilde H^\dag\ {l}_{1L})}{M}\
\left[2 \lambda_{1\nu} \lambda_{2\nu} + r\ \lambda_{3\nu}^2 \right] \left[\Sigma_{\alpha-1\,6}\Sigma_{6\,2+\alpha} \right]^{-2 Y_{1L}} \nn \\
& + &
 \frac{ (\overline{{l}^c_{1L} }\ \tilde H^*)(\tilde H^\dag\ {l}_{\alpha L}) + (\overline{{l}^c_{\alpha L} }\ \tilde H^*)(\tilde H^\dag\ {l}_{1L}) }{M}
 \lambda_{2\nu} (\lambda_{1\nu}' \ \epsilon_{\alpha\beta} \Sigma_{\beta-1\,6}
+ \lambda_{1\nu}''\ \Sigma_{6\,2+\alpha})
  \left[\Sigma_{\delta-1\,6}\Sigma_{6\,2+\delta} \right]^{-Y_{1L}+Y_{\nu_{2R}}}   \nn \\
&+ &
\frac{(\overline{{l}^c_{\alpha L} }\ \tilde H^*)(\tilde H^\dag\ {l}_{\beta L})}{2 M_3}
(i\sigma_2 \sigma_{\tau})_{\alpha\, \beta}\  (i\sigma_2 \sigma_{\tau})_{\delta\,\gamma}
\left[(\lambda_{3\nu}')^2 \ \Sigma_{\delta-1\,3}\Sigma_{\gamma-1 \,3}
\right.  \nn \\
&+ &  \left.
 \lambda_{3\nu}' \lambda_{3\nu}'' \ ( \epsilon_{\gamma\gamma'} \Sigma_{\delta-1\,3}\Sigma_{3\,2+\gamma'} + \delta\leftrightarrow\gamma )
+(\lambda_{3\nu}'')^2\ \epsilon_{\delta\delta'}\epsilon_{\gamma\gamma'}\Sigma_{3\,2+\delta'} \Sigma_{3\,2+\gamma'} \right] + H.c. \, ,
\eea
where $r = M/M_3$, $M$ and $M_3$ are the masses of the right-handed neutrinos, $\sigma_{\tau}/2$ are the generators of  the
$SU(2)_f$ gauge group ($\tau=1,2,3$).

The lagrangian for the charged leptons is given by
\bea \mathcal{L}_{e}
&=& \overline{e_{R}}\ \left[ \lambda_{1e}\
(\Sigma_{\alpha-1\,6}\Sigma_{6\,2+\alpha})^{(-Y_{1L}+Y_{1R})}\
(H^\dag\ {l}_{1L})
\right.  \nn \\
&& \quad \quad
+ \left.
i\ (\lambda_{3e}\  \Sigma_{6\,2+\alpha} + \lambda_{2e}\  \epsilon_{\alpha\,\beta} \Sigma_{\beta-1\,6})(\Sigma_{\delta-1\,6}\Sigma_{6\,2+\delta})^{(Y_{1R}-1)}\
(H^\dag\ {l}_{\alpha L}) \right]  \nn \\
&+&  \overline{E_{\alpha R}}\Big[ i
\left( \lambda_{1E}'\  \Sigma_{6\,2+\alpha} + \lambda_{1E}\ \epsilon_{\alpha\,\beta} \Sigma_{\beta -1\,6} \right)
(\Sigma_{\delta-1\,6}\Sigma_{6\,2+\delta})^{-Y_{1L}} \Big]
(H^\dag\ {l}_{1L})  \nn \\
&+&  \overline{E_{\alpha R}}\Big[
\delta_{\alpha\,\beta} \big(
- \lambda_{2E}\ + \lambda_{2E}'\ \Sigma_{\gamma-1\,6}\Sigma_{3\,2+\gamma}
+ \lambda_{2E}''\  \Sigma_{\gamma-1\,3}\Sigma_{6\,2+\gamma} \big)
  \nn \\
&& \quad\quad\
+ \Big(\lambda_{3E}\  \Sigma_{\alpha-1\,6}\Sigma_{3\,2+\beta}
+ \lambda_{3E}' \ \epsilon_{\alpha\,\delta}\epsilon_{\beta\,\gamma}\ \Sigma_{6\,2+\delta}\Sigma_{\gamma-1\,3} + (3\leftrightarrow 6) \Big) \nn \\
&& \quad\quad\
+\Big( \lambda_{4E}\  \Sigma_{\alpha-1\,6}\Sigma_{\gamma-1\,3}
\epsilon_{\beta\,\gamma}
+ \lambda_{4E}'\ \epsilon_{\alpha\,\delta}
\Sigma_{6, 2+\delta}\Sigma_{3, 2+\alpha}+ (3\leftrightarrow 6)
\Big)
\Big] (H^\dag\ {l}_{\beta L})\  + H.c. \, .
\label{yuklep}
\eea

\subsection{Leading order textures and masses}

On the vacuum that completely breaks the
the  $SU(2)_F\times U(1)_F$ gauge symmtery,
the little flavons acquire expectation values
$v_{1,2} = \varepsilon_{1,2} f$,
with $\varepsilon_{1,2} < 1$.
By inspection of the Yukawa lagrangians introduced in the previous section, we can determine the fermion  mass matrices.

Since all quarks are $SU(2)_F$ singlets the all entries of their mass matrices are proportional to powers of  $k\equiv \varepsilon_1 \varepsilon_2$.
Due to the large number of possible higher-order terms,  we only take for each entries the first non-vanishing term, and  obtain
\be
{M}^{(u)} = \langle h_0  \rangle
\left( \begin{array}{c c c}
 \lambda_{11} k^6 &  \lambda_{12} k^5 & \lambda_{13}  k^3 \\
 \lambda_{21} k^4 &  \lambda_{22} k^3  & \lambda_{23} k   \\
 \lambda_{31} k^3 &  \lambda_{32} k^2 &  \lambda_{33}
\end{array} \right)
\label{mass-up0}
\ee
and
\be
{M}^{(d)} = \langle h_0  \rangle \, k^2 \,
\left( \begin{array}{c c c}
\tilde \lambda_{11} k^5 &  \tilde \lambda_{12} k^4 & \tilde  \lambda_{13}k^2 \\
\tilde  \lambda_{21}k^3 & \tilde  \lambda_{22} k^2  & \tilde \lambda_{23}   \\
\tilde  \lambda_{31}k^3 & \tilde  \lambda_{32} k^2 & \tilde  \lambda_{33}
\end{array} \right) \, .
\label{mass-down0}
\ee

The essential feature of the previous mass matrices is that the fundamental textures
are determined by the vacuum structure alone---that is that obtained
by taking all Yukawa couplings $\lambda_{ij}$ and  $\tilde \lambda_{ij}$  of $O(1)$.
In fact, by computing the corresponding CKM matrix
one finds in first approximation
\be
V_{\rm CKM} =
\left( \begin{array}{c c c}
 1 &   O(k) & O(k^3) \\
 O(k) & 1 &   O(k^2) \\
  O(k^3)& O(k^2) &1
\end{array} \right) \, , \label{ckm0}
\ee
that is roughly of the correct form and, moreover,  suggests a value of $k\simeq \sin\theta_C \simeq 0.2$.

At the same time it is possible to extract from (\ref{mass-up0}) and (\ref{mass-down0}) approximated
 mass ratios:
\be
 \frac{m_u}{m_c} \simeq   \frac{m_c}{m_t}\simeq
\frac{m_d}{m_s} \simeq O(k^3) \quad   \frac{m_s}{m_b}\simeq O(k^2)
\ee
which again roughly agree with the experimental values.

These results show that the quark masses and mixing angles can be reproduced by our textures. While a rough agreement is already obtained by taking alla Yukawa coupling to be equal, the precise agreement with the experimental data depends on the actual choice of the Yukawa couplings $\lambda_{ij}$ and  $\tilde \lambda_{ij}$. However, their values can be taken  all of the same order, as we shall see in last section.

Notice that the textures used in this work do not satisfactorily address the
flavor problem in a supersymmetric framework:
the abelian nature of the flavor symmetry in the quark sector, and the
large mixing angles in the right handed mixing matrices $R^{(i)}$ would in general induce large contributions to FCNC processes via diagram with gluino exchange.
The diagonal entries of the squark mass matrices are not forbidden
by the abelian symmetry, and in general one expects all of them to be
determined only by the  scale of supersymmetry breaking,
up to ${\cal O}(1)$ coefficients.
Once fermions are diagonalized, large off-diagonal
entries are generated in  the $3\times 3$ right-handed down-type squark mass
matrix, because of the large mixing angles in $R^{(d)}$ (this can be easily seen from the fact that
second and third row of~\eq{mass-down0} have entries of the same order).
Phenomenologically, for generic choices of the diagonal elements of the
squark mass matrices, this leads to contributions to
$\Delta F=2$ processes ($K^0$-$\bar{K}^0$ or  $B^0$-$\bar{B}^0$ mixings and related CP violating observables) largely in excess of the experimental
data~\cite{GGMS}.
This can be avoided allowing for a degeneracy of the diagonal entries themselves, albeit with a tuning at least at the percent level.

In the lepton sector,  the VEV's of the little flavons gives us  the left-handed neutrino and charged-lepton mass matrices (again, we only retain the first non-vanishing term for each entry):
\begin{widetext}
\be
{ M^{(\nu)}} =  \frac{ \langle h_0 \rangle^2}{M}
\left( \begin{array}{c c c}    \left[ r\ \lambda_{3\nu}^2 + 2 \lambda_{1\nu} \lambda_{2\nu} \right]   \varepsilon_{1}^4\varepsilon_{2}^4
& -\lambda_{2\nu} \lambda_{1\nu}' \varepsilon_{1}^{2} \varepsilon_{2}  & -\lambda_{2\nu}\lambda_{1\nu}'' \varepsilon_{1} \varepsilon_{2}^{2} \\
-\lambda_{2\nu} \lambda_{1\nu}'   \varepsilon_{1}^{2} \varepsilon_{2}
& r\ \lambda_{3\nu}'^2 \varepsilon_{2}^2
& r\ \lambda_{3\nu}' \lambda_{3\nu}'' \varepsilon_{1} \varepsilon_{2}   \\
-\lambda_{2\nu} \lambda_{1\nu}'' \varepsilon_{1} \varepsilon_{2}^{2}
&  r\ \lambda_{3\nu}' \lambda_{3\nu}''  \varepsilon_{1} \varepsilon_{2}
&r\ \lambda_{3\nu}''^2 \varepsilon_{1}^{2}  \\
\end{array} \right)
\label{neutral-mass} \, .
\ee
\end{widetext}
The eigenvalues of this matrix are the masses of the three neutrinos.
The scale $M$  is just below or around $f$ and therefore we are not implementing the
usual see-saw mechanism that requires scales  as large as $10^{13}$ TeV. Therefore, realistic neutrino masses are obtained by tuning
the corresponding effective Yukawa couplings to the order of  $10^{-4}$
(which are the smallest couplings in the model).

In the same approximation, the Dirac mass matrix for the charged leptons is given by
\begin{widetext}
 \be
{ M}^{(l)} =   \langle h_0 \rangle
\left( \begin{array}{c c c}  \lambda_{1e}\, \varepsilon_{1}^4\varepsilon_{2}^4 & \lambda_{2e}\, \varepsilon_{1}^2 \varepsilon_2 & \lambda_{3e} \, \varepsilon_1 \varepsilon_{2}^2  \\
 \lambda_{1E} \,\varepsilon_{1}^{2} \varepsilon_{2}^3 & \lambda_{2E}  &
 (\lambda_{14E}^\prime + \lambda_{24E}^\prime) \,\varepsilon_{1}\varepsilon_{2}   \\
 \lambda_{1E}' \,\varepsilon_{1}^{3} \varepsilon_{2}^{2} &
 - (\lambda_{14E} + \lambda_{24E}) \,\varepsilon_{1}\varepsilon_{2}  &
 \lambda_{2E}   \\
\end{array} \right)
\label{charged-mass} \, ,
\ee
\end{widetext}
where the notation follows that of eq.~(39) in ref.~\cite{BBFP}.

In order to exhibit the main features of the underlaying textures, we study
the limit
\be
\varepsilon_1 \to 1\quad \mbox{and} \quad \varepsilon_2 \to k \, , \label{limit}
\ee
which is suggested by
the additional constraint
$  \varepsilon_1\varepsilon_2 \simeq \sin\theta_C$,
obtained from the study of the
quark textures.

Notice that in ref. \cite{BBFP} we have considered a slightly different charged-lepton
texture that accounts for maximal mixing in the  limit
$ \varepsilon_1^2 \ll \varepsilon_2^2 \ll 1$ (or, equivalently
$ \varepsilon_2^2 \ll \varepsilon_1^2 \ll 1$).

In the limit (\ref{limit}), the matrices in \eqs{neutral-mass}{charged-mass}  reduce---at the order $O(k^2)$, and up to overall factors---to
\be
M^{(\nu)} =
\left(\begin{array}{ccc}
0 & O(k) & O(k^2)\cr
O(k) & O(k^2) & O(k) \cr
O(k^2) & O(k)  & 1
\end{array}\right)
\quad \mbox{and} \quad
M^{(l)} =\left(\begin{array}{ccc}
0 & O(k) & O(k^2)\cr
0 & 1 & O(k)\cr
O(k^2) &O(k)& 1
\end{array}\right)\,, \label{m}
\ee
where, as before, the 1 stands for $O(1)$ coefficients.

The eigenvalues of $M^{(l)}$ can be computed by diagonalizing $M^{(l)\,\dagger}\,M^{(l)}$.
This product is---again for each entry to leading order in $k$:
\be
M^{(l)\,\dagger}\,M^{(l)}\,=\,\left(
\begin{array}{ccc}
0 & 0 & O(k^2)\cr
0 & 1 & O(k)\cr
O(k^2) & O(k) & 1
\end{array}
\right)\,.
\label{mm}
\ee

By inspection of the $2\times 2$ sub-blocks, the matrix \eq{mm} is
diagonalized by three rotations with angles, respectively,
$\theta_{23}^l \simeq \pi/4$ and $\theta_{12}^l\simeq \theta_{13}^l \ll 1$,
leading to one maximal mixing angle and two minimal. On the other
hand, the neutrino mass matrix in \eq{m} is diagonalized by three rotations
with angles, rispectively, $\tan 2\theta_{12}^\nu \simeq 2/k$ and
$\theta_{23}^\nu\simeq \theta_{13}^\nu \ll 1$
(the label $3$ denotes the heaviest eigenstate).
Therefore, the textures in the mass matrices in \eqs{neutral-mass}{charged-mass}
give rise to a PMNS mixing matrix---that is the combination of the
the two rotations above---in which $\theta_{23}$ is maximal,
$\theta_{12}$ is large (up to maximal), while $\theta_{13}$ remains small.

The natural prediction when taking  all  coefficients $O(1)$ is then:
a large atmospheric mixing
angle $\theta_{23}$, possibly maximal, another large solar mixing angle $\theta_{12}$, and a small $\theta_{13}$ mixing angle; at the same time,
the mass spectrum includes one light ($O(k^4)$)
and two heavy states ($O(1)$) in the charged
lepton sector ($m_e$, $m_\mu$ and $m_\tau$ respectively),
two light states ($O(k^2)$) and one heavy ($O(1)$) in the neutrino sector, thus predicting a neutrino spectrum with normal hierarchy.

By flavor symmetry, one expects masses of the same order
of magnitude for $\mu$ and $\tau$.
The ratio of the  masses of $\tau$ and
$\mu$  is given to $O(k)$ by:
\be
R\,\equiv\,\frac{m_{\mu}}{m_{\tau}}\, \simeq \,
\frac{\sqrt{\det\,(m^{(l)\,\dagger}\,m^{(l)})}}{\Tr\,(m^{(l)\,\dagger}\,m^{(l)})}\,,
\ee
where $m^{(l)}$ is the  $\mu$-$\tau$ sub-matrix of $M^{(l)}$.
The experimental splitting can be explained only
admitting a moderate amount of fine-tuning, of a factor 10, between
the coefficients of the charge lepton mass matrix
such as to make $R \simeq O(10^{-1})$.
One can quantify the stability of this fine-tuning
with the logarithmic derivatives $d^{R}_{Y_{ij}}$ of this ratio
with respect to the corresponding
Yukawa coefficients $Y_{ij}$~\cite{finetuning}:
\bea
d^R_{Y_{ij}}\,\equiv\,
\left|\frac{Y_{ij}}{R}\,\frac{\partial\,R}{\partial\,Y_{ij}}\right|\,.
\eea
Using the experimental value $R = m_{\mu}/m_{\tau}$, and the numerical solution
given in sect.\ III.A, we find ($i,j=2,3$)
$d^{R}_{Y_{ij}} < 5$,
where the largest value arises because of the leading order
correlation between the
diagonal Yukawas entries ($Y_{22}=Y_{33}$) in the charge lepton sector
that doubles the sensitivity.
In the absence of any fine-tuning one
would expect values of $d^R_{Y_{ij}}$ at most around unity.
Nevertheless, the tree level value of $R$
is not destabilized by Yukawa radiative corrections,
since they are very suppressed in the model.

The ultraviolet completion of the theory, in which all effective couplings
should be computed from a restrict number of fundamental parameters,
might explain possible correlations among the
Yukawa couplings, together with the
suppression of the the overall neutrino scale.

Finally, notice that even though the quark charges leave an undetermined factor $q$ (see Table I), gauge anomalies are  present in the theory, as it can be easily seen by inspection considering the charges of the matter fields. They can be cancelled by adding appropriate Wess-Zumino terms~\cite{FPPS}.

\section{Fitting the Data}

Let us first briefly review the experimental data and comment on the possible range of values we consider acceptable in reproducing these data within the model.

The CKM matrix is rather well known as are the masses of the quarks (see, e.g., the PDG~\cite{PDG}). We will estimate only ratios of masses which are
renormalization group invariant, so that we only have to be careful in computing them at a common scale.
Taking into account the uncertainties in the values of the quark masses,  the mass ratio we would like the model to reproduce are given by
\be
\frac{m_t}{m_c} = 248 \pm 70 \quad \frac{m_b}{m_s} = 40 \pm 10 \quad
\frac{m_s}{m_d} = 430 \pm 300 \quad \frac{m_c}{m_u} = 325 \pm 200  \, .
\ee

The CKM phase is determined~\cite{delta} to be
\be
\delta  = 61.5^o \pm 7^o \quad
(\sin 2\beta= 0.705 ^{+0.042}_{-0.032}) \, .
\label{expKMphase}
\ee

Compelling evidences in favor of neutrino oscillations and, accordingly of non-vanishing neutrino masses has been collected in recent years from neutrino experiments~\cite{experiments}. Combined analysis of the experimental data show that the neutrino mass matrix is characterized by a hierarchy with two square mass differences (at $99.73\%$ CL):
\bea
\Delta m^2_\odot & = & (5.3 - 17) \times 10^{-5} \mbox{eV}^2 \nn\\
|\Delta m^2_\oplus | &= & (1.4 - 3.7) \times 10^{-3} \mbox{eV}^2 \, , \label{delta-masses}
\eea
the former controlling solar neutrino oscillations~\cite{Sandhya} and the latter the atmospheric neutrino experiments~\cite{Fogli}.
In the context of three active neutrino oscillations, the mixing
is described by the
PMNS mixing matrix $V_{PMNS}$ in \eq{pmns}.
Such a matrix is parameterized by three mixing angles, two of which
($\theta_{12}$ and $\theta_{23}$) can be identified with the mixing angles
determining solar~\cite{Sandhya} and atmospheric~\cite{Fogli} oscillations,
respectively (again, at $99.73\%$ CL):
 \bea
 \tan^2 \theta_\odot & = &  0.23 - 0.69 \, ,  \nn \\
 \sin^2 2\, \theta_\oplus  & = & 0.8 - 1.0 \, .  \label{mix-angles}
 \eea

For the third angle, controlling the mixing $\nu_\tau$-$\nu_e$, there are at present only upper limits, deduced by reactor neutrino experiments~\cite{reactors} (at 99.73\% CL):
\be
\sin^2  \theta_{13} < 0.09 \, .
\ee

Other observable quantities determined by the neutrino mass matrix have
not been measured yet. These include: 1) the type of neutrino
spectrum, with normal or inverted hierarchy (see for instance~\cite{CPP} for a definition),
2) the common mass scale, i.e. the actual value of the lowest mass eigenvalue $m_1$,
3) the (Dirac) phase $\delta^{l}$ responsible for CP violation in leptonic
flavor changing processes,
4)  the two Majorana flavor-diagonal CP-violating phases,
5) the sign of $\cos2\theta_{\oplus}$.
Several proposal appeared in the literature
to measure all these quantities in the next generation neutrino experiments,
together with the mixing angle $\theta$~\cite{future}.
Our model predicts a neutrino spectrum with normal hierarchy,
with a very small mass for the lighter neutrinos
$m_{1,2}\ll \sqrt{\Delta m^2_{\oplus}}$.

Finally, the values of the charged-lepton masses are given by
$m_\tau\simeq 1777$ MeV, $m_\mu\simeq 106$ MeV and  $m_e
 \simeq 0.51$ MeV, respectively. We therefore have
 \be
 \frac{m_\tau}{m_\mu} \simeq 17\, , \quad
\frac{m_\mu}{m_e} \simeq  207 \, , \quad
\frac{m_\tau}{m_e} \simeq  3484 \, .
\ee

\subsection{Masses and mixings}

In order to show that the model reproduces in a natural manner all the
experimental data we retain the first non-vanishing contribution to
each entry in all mass matrices and then---having extracted an overall
coefficient for each matrix according to \eqs{mass-up0}{mass-down0}
and \eqs{neutral-mass}{charged-mass}---treat the ratios of Yukawa
couplings as a set of arbitrary parameters to be varied within a
$O(1)$ range.

We keep the VEV's $v_1$ and $v_2$ fixed at the values   obtained by taking $\varepsilon_1=0.8$ and $\varepsilon_2=0.2$.

In practice, we generated for the quark matrices many sets of 18 complex Yukawa
parameters whose moduli differ by at most a factor 10
and accepted those that reproduces the known masses and mixings.
As an example, we found that the assignments
\be
\left[ \begin{array}{c c c}
 \lambda_{11}  &  \lambda_{12}  & \lambda_{13}  \\
 \lambda_{21} &  \lambda_{22}   & \lambda_{23}    \\
 \lambda_{31}  &  \lambda_{32} &  \lambda_{33}
\end{array} \right] =  \lambda_U\; \left[ \begin{array}{c c c}
1.2 + 0.073 i & 1.9 +0.31 i  & -0.82+1.3 i \\
 -0.32 -0.41 i & -0.58 +0.85 i   & -0.48 -0.95 i \\
1.2 + 0.84 i &  -1.5 +0.78 i &  1.4 + 0.72 i
\end{array} \right]
\label{up-fit}
\ee
and
\be
\left[ \begin{array}{c c c}
\tilde \lambda_{11} &  \tilde \lambda_{12}  & \tilde  \lambda_{13}\\
\tilde  \lambda_{21}& \tilde  \lambda_{22}  & \tilde \lambda_{23}   \\
\tilde  \lambda_{31} & \tilde  \lambda_{32} & \tilde  \lambda_{33}
\end{array} \right] =
\lambda_D\;  \left[ \begin{array}{c c c}
-0.55 - 1.5 i & -0.76 -0.42 i  &  0.55+1.2 i \\
 -1.3 -0.83 i & 0.32 +1.2 i   &  0.58 +0.67 i \\
 0.75 -1.0 i &  -1.4 +0.17 i &  0.09 -1.6 i
\end{array} \right]
\label{down-fit}
\ee
with $\lambda_U$ and $\lambda_D$ of $O(1)$,
give masses and mixing angles in excellent
agreement with the experimental data.
We have followed a similar procedure for the leptonic sector,  generating
random sets of 13 real parameters.
Lacking experimental signature
of CP violation in the leptonic sector, we have neglected,
for the purpose of illustration,
leptonic phases in the numerical exercise.
Again, we obtain that for the representative choice
\begin{widetext}
\be
 \left[ \begin{array}{c c c}    r\ \lambda_{3\nu}^2 + 2 \lambda_{1\nu} \lambda_{2\nu}   & -\lambda_{2\nu} \lambda_{1\nu}'  & -\lambda_{2\nu}\lambda_{1\nu}''  \\
-\lambda_{2\nu} \lambda_{1\nu}'
& r\ \lambda_{3\nu}'^2 & r\ \lambda_{3\nu}' \lambda_{3\nu}''  \\
-\lambda_{2\nu} \lambda_{1\nu}''
&  r\ \lambda_{3\nu}' \lambda_{3\nu}'' &r\ \lambda_{3\nu}''^2 \\
\end{array} \right] =
\lambda_\nu^2\; \left[ \begin{array}{c c c}
0.66 &  -1.0&  2.9\\
-1.0 & 1.9  &  0.29 \\
2.9 & 0.29 & -1.1
\end{array} \right]
\ee
\end{widetext}
with $\lambda_\nu = O(10^{-4})$, and
\begin{widetext}
\be
\left[ \begin{array}{c c c}  \lambda_{1e} & \lambda_{2e}& \lambda_{3e}  \\
 \lambda_{1E} & \lambda_{2E}  & \lambda_{4E}'  \\
 \lambda_{1E}' & - \lambda_{4E}& \lambda_{2E}  \\
\end{array} \right] =  \lambda_E\; \left[ \begin{array}{c c c}
 1.2& 0.27 & 1.4  \\
 -1.2 & 0.39 & 2.3 \\
 0.36& 2.0 &  0.39
\end{array} \right]
\ee
\end{widetext}
with $\lambda_E = O(10^{-2})$,
the experimental values are well reproduced.
\begin{table}[ht]
\begin{center}
\caption{Experimental data vs.\ the result of our numerical analysis based on a representative set of Yukawa couplings of order one (see text)
and  $\varepsilon_1=0.8$ and $\varepsilon_2=0.2$.
Uncertainties in the experimental inputs are explained in the main body.}
\label{data}
\vspace{0.2cm}
\begin{tabular}{|c|c|c|}
\hline
\null & exp & numerical results \cr
\hline
\hline
$|V_{us}|$ &   $0.219-0.226$ & 0.22 \cr
$|V_{ub}|$  & $0.002-0.005$ &  0.0035\cr
$|V_{cb}|$  & $0.037-0.043$ &   0.040\cr
$|V_{td}|$  & $0.004-0.014$ &  0.0079\cr
$|V_{ts}|$  & $0.035-0.043$ &   0.039\cr
$\delta$& $61.5^{o}\pm 7^{o}$  & $61^{o}$ \cr
$\sin 2\beta $ & $0.705^{+0.042}_{-0.032}$ & 0.69\cr
\hline
$m_t/m_c$  & $248\pm 70$ & 219  \cr
$m_c/m_u$ & $325\pm 200$ & 300 \cr
$m_b/m_s$  & $40\pm 10$ &  45 \cr
$m_s/m_d$  & $430\pm 300$ & 231  \cr
\hline
\hline
$\tan^2 \theta_\odot$&$0.23-0.69$ &0.32 \cr
$\sin^2 2\theta_\oplus$ & $0.8-1.0$ & 1.0\cr
$\sin^2 \theta_{13}$ & $<0.09$& 0.08 \cr
\hline
$\Delta m^2_\odot /\Delta m^2_\oplus $&$0.014-0.12$  &0.043 \cr
$m_\tau/m_\mu$& $17$ & 15\cr
$m_\mu/m_e$& $207$ & 231\cr
$m_\tau/m_e$ & $3484$ & 3465 \\[1ex]
\hline
\end{tabular}
\end{center}
\end{table}

Table III summarizes the experimental data and compares them to the result of the above procedure. The agreement is quite impressive, keeping in mind that
we have varied only the leading terms in the mass matrices.
While the values of the overall constants (which are related to the scale
of the heaviest state in the mass matrices)
are not explained by
the model, the hierarchy among the mass eigenvalues and the
mixing angles are given in first approximation by the flavor symmetry
and the flavor vacuum so that, within each sector,
the Yukawa couplings remain in a natural range.

The phenomenology related to the gauge boson and flavon
couplings with quark and leptons will be studied in a forthcoming work where direct bounds for the masses of the flavons will be derived. We expect 
these to come mainly from flavor gauge mediated processes
since the flavon effective couplings to matter
are suppressed by powers of weak and flavor vacuum expectations values
over the flavon scale $f$.


 \acknowledgments

 MF and MP thank the Aspen Center for Physics
for the
 hospitality during part of the work. MP thanks S.\ Vempati for useful discussions.
MP thanks also the Harvard High Energy Theory Group  and the
Theoretical Physics Group of the Lawrence Berkeley National Laboratory,
for hospitality during the completion of this research.
The work of FB, SB and MF
is partially supported by the European TMR Networks
HPRN-CT-2000-00148
 and HPRN-CT-2000-00152. The work of MP is
supported in part by the
 US Department of Energy under contract
DE-FG02-92ER-40704.



 \end{document}